\documentclass[12pt,a4paper,final]{iopart}

\usepackage{iopams}  
\expandafter\let\csname equation*\endcsname\relax
\expandafter\let\csname endequation*\endcsname\relax
\usepackage{amsmath}
\usepackage{graphicx}

\usepackage[breaklinks=true,colorlinks=true,linkcolor=blue,urlcolor=blue,citecolor=blue]{hyperref}

\begin{document}

\title[]{Finding Non-Zero Stable Fixed Points of the Weighted Kuramoto Model is NP-hard}

\author{Richard Taylor}
\address{Defence Science and Technology Organisation (DSTO), ACT 2600, Australia}
\ead{richard.taylor@dsto.defence.gov.au}
\begin{abstract}
The Kuramoto model when considered over the full space of phase angles [$0,2\pi$) can have multiple stable fixed points which form basins of attraction in the solution space. In this paper we illustrate the fundamentally complex relationship between the network topology and the solution space by showing that determining the possibility of multiple stable fixed points from the network topology is NP-hard for the weighted Kuramoto Model. In the case of the unweighted model this problem is shown to be at least as difficult as a number partition problem, which we conjecture to be NP-hard. We conclude that it is unlikely that stable fixed points of the Kuramoto model can be characterized in terms of easily computable network invariants.
\end{abstract}
%
%
\section{Introduction}
\label{intro}
The Kuramoto model \cite{Kuramoto1975} was originally motivated by the phenomenon of collective synchronisation whereby a system of coupled oscillating nodes will sometimes lock on to a common frequency despite differences in the natural frequencies of the individual vertices. This model has since been applied to a wide variety of application fields including, biological, chemical, engineering, and social systems (see the survey articles including \cite{Acebron2005}, \cite{Arenas2008}, \cite{Strogatz2000}, and recently \cite{Dorfler2013}). While Kuramoto studied the infinite complete network it is natural to consider finite networks of any given topology. This would correspond to a notion of coupling that is not universal across all node pairs, but rather applies to a subset of all possible links. For example the work patterns of human individuals in an organisation might enjoy a coupling effect in relation to pairs of individuals that have a working relationship. In this article we explain the significance of stable fixed point solutions to the model, and explore the algorithmic complexity of finding such solutions.
\subsection{The model}
Each node has an associated phase angle $\theta_i$, as well as its own natural frequency $\omega_i$.  The basic governing equation is the differential equation:
\begin{equation}
\dot{\theta}_i = \omega_i + k \sum_{j=1}^n A_{ij} \sin (\theta_j - \theta_i),   i=1,..,n.
\end{equation}
where $A$ is the adjacency matrix of the network and $k$ is a coupling constant which determines the strength of the coupling.  Note that each $\theta_i$ is understood to denote a function of time $t$. An important special case is where all the natural frequencies $\omega_i$ are equal. We shall call this case \textit{homogeneous} and the general case as \textit{inhomogeneous}. Typically $A$ is a (0,1)-matrix, however it may also be any real positive symmetric matrix reflecting coupling constants that vary for each edge. This case is of particular interest to us and we specify this by referring to it as the \textit{weighted} Kuramoto model. It is well known that results for unique stable fixed points can be obtained for restricted phase angles $\theta_i$, say in the range $[0,\pi]$. In this article we are concerned with results about stable fixed points over the full range $[0,2\pi)$.
\subsection{Synchronisation}
It has been observed that for arbitrary initial phases, networks \textit{synchronize} in that some of the node phases converge to the same, or nearly the same phase angle, while the phase frequencies $\dot{\theta}_i$ converge to a common value. Meanwhile, the remaining nodes behave non-uniformly or 'drift'. Moreover, at a sufficiently large \textit{critical} coupling constant $k$, it becomes possible for all nodes to participate in synchronized behaviour. By summing over equation (1), the sine terms cancel and the average frequency of the nodes is a constant $\bar{\omega}$ so that
\begin{equation}
\frac{1}{n}\sum_{j=1}^n\dot{\theta}_j=\frac{1}{n}\sum_{j=1}^n\omega_j= \bar{\omega}.
\end{equation}
Define a \textit{frequency fixed point} as a situation in which all the node frequencies are equal and are fixed over time. By equation (2), this is characterized by
\begin{equation}
\dot{\theta}_i=\bar{\omega},   i=1,..,n.
\end{equation}
It follows that for a frequency fixed point all phase angles remain constant. We note that a frequency fixed point may also be accompanied by \textit{phase synchrony} whereby the nodes have the same, or nearly the same, phase angles. This is however not necessarily the case, and the phase angles of a frequency fixed point may be significantly different. Examples of this kind are the ring networks \cite{Ochab2010}. 
Combining equations (1) and (3), the phase differences at a frequency fixed point satisfy
\begin{equation}
\bar{\omega} = \omega_i + k \sum_{j=1}^n A_{ij} \sin (\theta_j - \theta_i),   i=1,..,n.
\end{equation}
Henceforth, we shall use the term fixed point to refer to a frequency fixed point satisfying equation (4). The  character, number and location of the fixed points of a network are clearly important in understanding the types of dynamics that are possible from all possible initial phases of the system. In particular, understanding the nature of all stable fixed points is particularly important, since each represent an attractor set of positive $n$-dimensional volume within the set of all states. Fixed points can have a range of attractor types ranging from unstable single points, $m$-dimensional ($m<n$) saddle point structures that are partially stable, to $n$-dimensional volumes for stable fixed points.
\subsection{Stability}
Let $\{\theta^{*}_{i},i=1,..,n \}$  be a fixed point so that by equation (3)
\begin{equation}
\dot{\theta}^{*}_i(t)=\bar{\omega},   i=1,..,n.
\end{equation}
Since ${\theta}^{*}_{j}(t)-{\theta}^{*}_{i}(t)$ is a constant for all $i,j$ we use the compact notation
 ${\theta}^{*}_{j}(t)-{\theta}^{*}_{i}(t)={\Delta}^{*}_{ji}$. A linear stability analysis shows the fixed point to be stable (see for eg. \cite{Taylor2012}) if and only if the matrix $M$ is negative definite where
\begin{equation}
M_{ij}=A_{ij}\cos({\Delta}^{*}_{ji})-{\delta}_{ij} \sum_{k=1}^n A_{ik}\cos({\Delta}^{*}_{ki}),
\end{equation}
and where the Kronecker delta function is defined as $\delta_{ij}=1,0$ for $i=j$ and $i\neq j$, respectively.
For a fixed point ${\theta}^{*}_i,i=1,..,n$, we can rearrange equation (4) as
\begin{equation}
\frac{\bar{\omega}-\omega_i}{k}= \sum_{j=1}^n A_{ij} \sin ({\Delta}^{*}_{ji}),   i=1,..,n.
\end{equation}
It follows that as the coupling constant increases in relation to the natural frequencies the left hand side of equation (7) tends to $0$, in the limit giving:
\begin{equation}
\sum_{j=1}^n A_{ij} \sin ({\Delta}^{*}_{ji})=0,   i=1,..,n.
\end{equation}
Note that this limiting equation is the same as that obtained in the homogeneous case, and also that this equation is independent of the value of the equal natural frequency. Thus as the coupling constant increases the fixed point for unequal natural frequencies converge to the solutions for the homogeneous system of equations (8). Furthermore, in the homogeneous case it is sufficient to consider the case where the natural frequencies are all zero. For this reason the fixed point character for the case where all natural frequencies are zero is fundamental to understanding the limiting behaviour of the general inhomogeneous case. Similarly the stability condition does not depend on the natural frequencies (see equation (6)). Thus the homogeneous case can in some sense be considered as a guage of the contribution of the purely topological effects (from $A_{ij}$) on fixed points and stability.
Though the relationship between the network topology $A_{ij}$ and the stable fixed points is a complex one, we have the following useful necessary condition \cite{Taylor2012}.
Let $\{\theta^{*}_i,i=1,..,n\}$ be any stable fixed point solution to the homogeneous Kuramoto model. Then for any proper node subset $X$ we have the equality
\begin{equation}
\sum_{(i,j)\in(X,X^c)} A_{ij} \sin ({\Delta}^{*}_{ji})=0.
\end{equation}
and inequality
\begin{equation}
\sum_{(i,j)\in(X,X^c)} A_{ij} \cos ({\Delta}^{*}_{ji}) > 0.
\end{equation}
It follows that if for every edge $(i,j)$ if $|{\Delta}^{*}_{ji}|>\pi/2$, then the fixed point ${\theta}^{*}_i,i=1,..,n$ is unstable. It is also well known \cite{Ochab2010} that if for each link $(i,j)$ $|{\Delta}^{*}_{ji}|<\pi/2$ then the fixed point is stable. These equations are useful in resolving the mixed case where some $|{\Delta}^{*}_{ji}|$ are less than $\pi/2$ and some are greater than $\pi/2$. We shall often find the case where the sets $X$ in (9) and (10) are single nodes particularly useful, where (9) and (10) take the form
\begin{eqnarray}
\sum_{j=1}^n A_{ij} \sin ({\Delta}^{*}_{ji})=0,   i=1,..,n.\\
\sum_{j=1}^n A_{ij} \cos ({\Delta}^{*}_{ji})>0,   i=1,..,n.
\end{eqnarray}
It is useful to visualize the phase angles as points on the unit circle within the complex plane where the nodes are positioned around a circle, which we shall refer to as a \textit{phase angle diagram}. Thus we let $z_{i}, i=1,..,n$ be given by $z_{j}=e^{i{\theta}^{*}_j}=\cos({\theta}^{*}_j)+i\sin({\theta}^{*}_j)$. This is illustrated in figure 1 with the representation of the hexagonal ring graph which forms a non-zero stable fixed point. There is also a simple physical intuitive way of understanding the phase angle diagram by interpreting each edge as an elastic link between beads on a frictionless ring. A fixed point corresponds to the elastic forces at each bead balancing out, and stability is manifested by any small perturbations of the beads returning to the fixed point (see \cite{Dorfler2013}).
\begin{figure}
\begin{center}
\includegraphics[width=60mm]{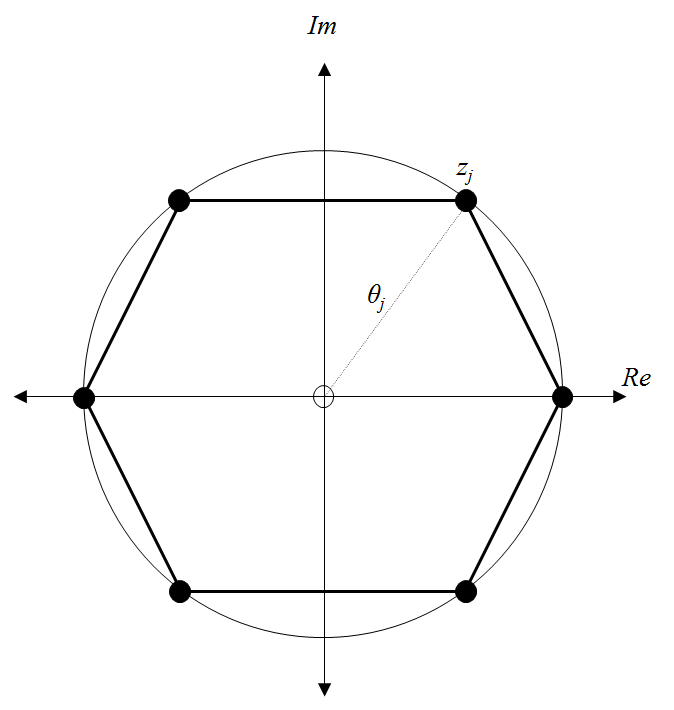}
\caption{\label{figure1.png} Representation of nodes on the complex plane}
\end{center}
\end{figure}
\section{Non-zero stable fixed points}
Though the nature and presence of stable fixed points is well understood for certain classes of graphs, general results for arbitrary graph topologies seem difficult to obtain. Thus complete graphs can have only one single stable fixed point \cite{Monzon2005} as can all dense graphs with degree at least $0.9395(n-1)$ \cite{Taylor2012}. Also trees can have only one stable fixed point \cite{Canale2010} though the nature of all fixed points and the relationship to the coupling constant is well understood \cite{Dekker2013}. On the other hand ring graphs do have multiple stable fixed points the nature of which is also well understood \cite{Ochab2010}. In contrast we show that the most basic of questions about the stable fixed points of the weighted Kuramoto model is unlikely to be computable efficiently. Specifically, determining the presence of any non-zero stable fixed points from the network topology $A_{ij}$ even for the homogeneous system is NP-hard. This means that an efficient (polynomial time) algorithm to solve this problem is either very difficult to find or does not exist (see the standard guide to algorithmic complexity \cite{Garey1979}). This remains true even if the phase angle differences across edges ($A_{ij}\neq 0$) are restricted to $|{\Delta}^{*}_{ji}|<\pi/2$.  For the unweighted case we show that determining the presence of any non-zero stable fixed points from the network topology $A_{ij}$ is at least as difficult (up to polynomial complexity) as solving a particular number problem. To establish the former we shall refer to the Partition problem  \cite{Garey1979}. In terms of related algorithmic complexity results we are aware only of the result of \cite{Rohn1994} which shows that checking positive definiteness of symmetric interval matrices is NP-hard. This problem is significantly different to the problem presented here in a number of ways however. For example general symmetric matrices have $\frac{1}{2}n(n-1)$ degrees of freedom wheras the Kuramoto graphs have $n-1$ degrees of freedom (without loss of generality any one node can be fixed at phase angle $0$). Also we seek negative definiteness in conjunction with $n$ equality conditions in the form of the frequency fixed point equations (8).
\subsection{The weighted Kuramoto model and the Partition problem}
Using the standard formalism of  \cite{Garey1979}\\
\textbf{Partition} \\
Instance: A positive integer $m$ and a collection of positive integers $a_{1},a_{2},..,a_{m}$. \\
Question: Is there a subset $S$ of $\{1,2,..,m\}$ in which
\begin{equation}
\sum_{i\in S} a_{i}=\sum_{i\in{S^c}} a_{i}?
\end{equation} 
For any collection of $m$ positive integers $a_{1},a_{2},..,a_{m}$ we construct a particular weighted graph $G[a_{1},a_{2},..,a_{m}]$, shown in figure 2, with corresponding adjacency matrix $A_{ij}$ and show that Partition is satisfied if and only if there is a  non-zero stable fixed point for the corresponding homogeneous Kuramoto model. 

Let  $t=1/2\sum_{i} a_{i}$. Set the vertex set of $G$ to be $V=x, u_{1}, u_{2},.., u_{m}, v_{1}, v_{2},.., v_{m}, y$. For each $i=1,..,m$ there is a link of weight $a_{i}$ between $x$ and $u_{i}$, a link of weight $a_{i}$ between $u_{i}$ and $v_{i}$, and a link of weight $2ta_{i}$ between $v_{i}$ and $y$.
\begin{figure}
\begin{center}
\includegraphics[width=60mm]{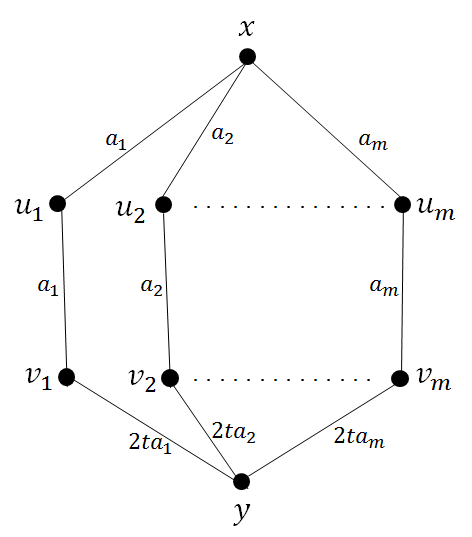}
\caption{\label{figure2.png} The graph $G[a_{1},a_{2},..,a_{m}]$ }
\end{center}
\end{figure}
Let \{${\theta}^{*}_i,i=1,..,n$\} be any non-zero stable fixed point of the homogeneous Kuramoto model corresponding to $A_{ij}$. We shall show that there is a  non-zero stable fixed point for $A_{ij}$ if and only if Partition is satisfied for inputs $m,a_{1},a_{2},..,a_{m}$. To do this we first assume that there is a subset $S$ satisfying equation (11). Consider the assignment of phase angles shown in the phase angle diagram figure 3 where the nodes are assigned from six different phase angles shown within the diagram at $0$, $\alpha$, $2\alpha$, $\pi$, $2\pi-2\alpha$, and  $2\pi-\alpha$. The open nodes refer to groups of nodes with the same phase angle. The total edge weights between node clusters are indicated by numbers on the links. It is straightforward to verify that this is a fixed point solution where $\alpha$=arcos$(1/2t)$. Since all the phase angle differences across edges are also between $(-\pi/2,\pi/2)$ the fixed point is also stable.
\begin{figure}
\begin{center}
\includegraphics[width=70mm]{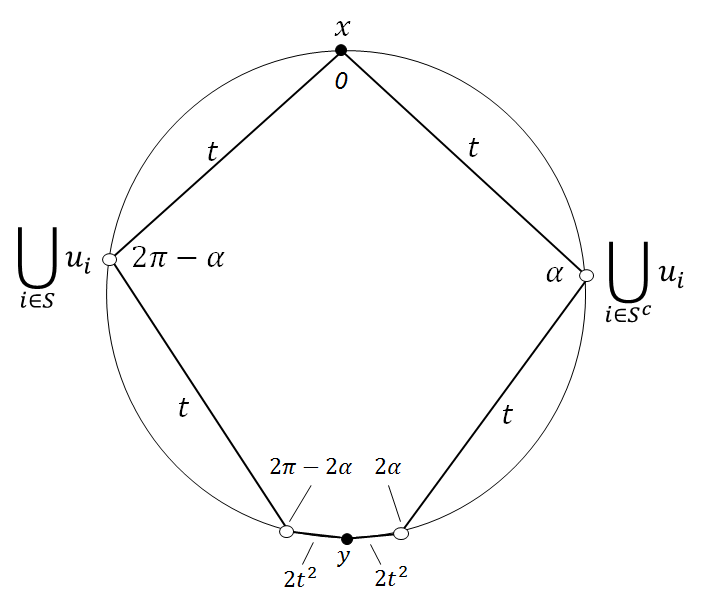}
\caption{\label{figure3.png} non-zero stable fixed point solution}
\end{center}
\end{figure}
Now assume that the system has a non-zero stable fixed point $\{\theta^{*}_i,i=1,..,n\}$ of the corresponding weighted homogeneous Kuramoto model. Let the stable fixed point angles corresponding to the nodes of $G$ be denoted by $\theta^*[x]=0$ at $x$, $\theta^*[u_{i}]$ at $u_{i}$, $\theta^*[v_{i}]$ at $v_{i}$ and  $\theta^*[y]$ at $y$. If for each $i$ we apply equation (8) to vertex $v_{i}$ we have
\begin{equation}
ta_{i}\sin(\theta^*[v_{i}]-\theta^*[y])=a_{i}\sin(\theta^*[u_{i}]-\theta^*[v_{i}])
\end{equation}
so that $|\sin(\theta^*[v_{i}]-\theta^*[y])| \leq 1/2t$. For $t$ large this means that $\theta^*[v_{i}]-\theta^*[y]$ is close to $0$ or $\pi$. However by the stability inequality (10) applied at $X=\{v_{i}\}$, 
\begin{equation}
ta_{i}\cos(\theta^*[v_{i}]-\theta^*[y])+a_{i}\cos(\theta^*[u_{i}]-\theta^*[v_{i}]) \geq 0. 
\end{equation}
This rules out the latter possibility (that $\theta^*[v_{i}]-\theta^*[y]$ is close to $\pi$)  and for $1/2t$ small certainly  $|\theta^*[v_{i}]-\theta^*[y]| \leq 1/t$. We must now have the phase angles arranged as in the phase angle diagram of figure 4 where $f$ and $g$ are integers with $f+g=\sum a_{i}$ and say $f \leq g$.
\begin{figure}
\begin{center}
\includegraphics[width=70mm]{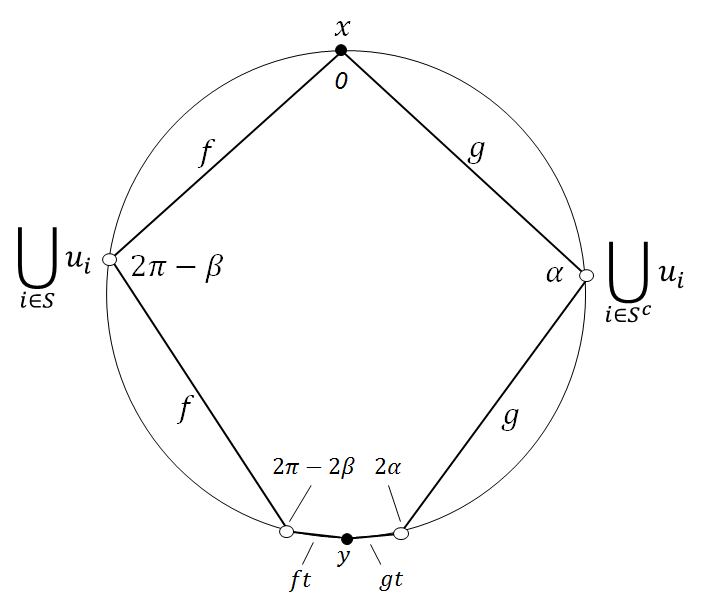}
\caption{\label{figure6.png} proving $t=u$ forming a partition of $a_{1},a_{2},..,a_{m}$}
\end{center}
\end{figure}
For stability $\beta \leq \pi/2$ so that $\alpha \geq \pi/2-1/t$. By equation (8) at $x$ 
\begin{equation}
f \geq g\sin(\frac{\pi}{2}-\frac{1}{t}) \geq g-\frac{g}{t}.
\end{equation}
Since $t>g$ and $f$ and $g$ are integers we must have $f=g$ and thus we have found a partition of $a_{1},a_{2},..,a_{m}$. Thus we have shown that Partition is satisfied for inputs $m,a_{1},a_{2},..,a_{m}$ if and only if there is a non-zero stable fixed point for the corresponding weighted homogeneous Kuramoto model. Since Partition is NP-complete \cite{Garey1979} it follows that determining whether weighted Kuramoto models have a non-zero stable fixed point is NP-hard. Note that it does not follow that this problem is NP-complete since it is not at all clear that this problem is in NP.
Establishing NP-hardness for the case where $A$ is a $\{0,1\}$ matrix seems to be more difficult. By replacing each vertex $u_{m}$ and $v_{m}$ with a clique of size $a_{m}$ we can obtain an unweighted graph $G'[a_{1},a_{2},..,a_{m}]$ corresponding to $G[a_{1},a_{2},..,a_{m}]$ with $G'$ having a non-zero stable fixed point if and only if $G$ does. However $G'$ and the corresponding matrix $A'_{ij}$ require $\sum a_{i}$ bits of data in their representation, while the Partition problem is known to have an algorithm that runs in time polynomial in $\sum a_{i}$ \cite{Garey1979}. What we seek is an unweighted graph $G$ such that the problem of detecting non-zero stable fixed points might require exponential time in the size of $G$ (or is NP-hard in a strong sense \cite{Garey1979}). To do this in the following we refer to a real valued variation of the Partition problem. 
\subsection{The 0-1 Kuramoto model and a partition problem over the reals}
The following describes a variant of the Partition problem over the real numbers which we call Kuramoto Partition. In the body of this section we shall construct an unweighted graph and corresponding $(0,1)$ adjacency matrix such that  Kuramoto Partition is satisfied if and only if there is a  non-zero stable fixed point for the corresponding unweighted homogeneous Kuramoto model.\\
\textbf{Kuramoto Partition}\\
Instance: A positive integer $n$ and collection of positive integers $b_{i}, i=1,..,n$ such that $2 \leq b_{1}\leq b_{2}\leq..\leq b_{n}\leq n$.  \\
Question: Let $b_{i}=n, i=n+1,..,3n$. Is there a subset $S$ of $\{1,2,..,3n\}$ in which
\begin{eqnarray}
\sum_{i\in S}  \sqrt{\frac{b_{i}^2 (1-\epsilon^2)}{1+b_{i}^2+2b_{i}\epsilon}} = \sum_{i\in S^c}  \sqrt{\frac{b_{i}^2 (1-\epsilon^2)}{1+b_{i}^2-2b_{i}\epsilon}} \\
\frac{1}{n}>\epsilon\geq 0?
\end{eqnarray} 
\\
We note by continuity that (17) and (18) have a solution if and only if 
\begin{eqnarray}
\sum_{i\in S}  \frac{b_{i}}{\sqrt{1+b_{i}^2}} \geq \sum_{i\in S^c}  \frac{b_{i}}{\sqrt{1+b_{i}^2}} \textrm{ and }\\
\sum_{i\in S}  \sqrt{\frac{b_{i}^2 (1-1/n^2)}{1+b_{i}^2+2b_{i}/n}} < \sum_{i\in S^c}  \sqrt{\frac{b_{i}^2 (1-1/n^2)}{1+b_{i}^2-2b_{i}/n}}.
\end{eqnarray} 
Notice that Kuramoto Partition has input size $nlog_{2}n$. We show that any polynomial algorithm for detecting non-zero stable fixed points in the (unweighted) homogeneous Kuramoto model must produce a polynomial algorithm for Kuramoto Partition. We briefly provide some background to this type of problem. The Partition problem over sets of general real numbers have good heuristic methods. For example the differencing method \cite{Karmarker1982} tries to find a partition of a set of real numbers between 0 and 1 that minimises the difference between the sum of terms in each part, and runs in time $O(n)$.  Probabilistic analysis \cite{Karmarker1986} shows that for independent random selections of the real numbers, the minimum difference has a median that is exponentially small, while the differencing method will find differences as small as $O(n^{-clogn}),  c>0$ with the exception of pathological or rare examples. This suggests that for most instances Kuramoto Partition is satisfied and that solutions can be found in polynomial time. On the other hand for some instances Kuramoto Partition is not satisfied (for example $b_{i}=n$ for all $i$ and $n$ is odd), and is less likely to be satisfied if the $b_{i}$ have few different values.  In summary despite good heuristic methods the author is not aware of any polynomial algorithm (in $n$) for this problem nor whether Kuramoto Partition is NP-hard. 

Now let $b_{i}, i=1,..,n$ be integers such that $2 \leq b_{1}\leq b_{2}\leq..\leq b_{n}\leq n$. Also let $b_{i}=n, i=n+1,..,3n$.  Construct a graph $G[b_{1},b_{2},..,b_{3n}]$ as follows. A vertex $x$ is adjacent to nodes $u_{1},..,u_{3n}$, and a vertex $y$ is adjacent to nodes $v_{1},..,v_{3n}$. Additionally for each $i=1,..,3n$ both $u_{i}$ and $v_{i}$ are adjacent to separate complete graphs on $b_{i}$ nodes. This is illustrated in figure 5. 
\begin{figure}
\begin{center}
\includegraphics[width=60mm]{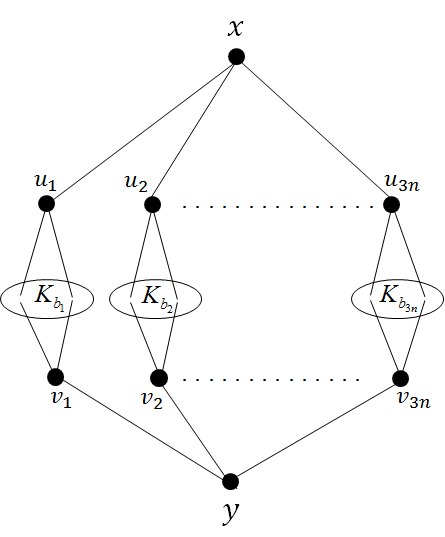}
\caption{\label{figure5.png} The graph $G[b_{1},b_{2},..,b_{3n}]$}
\end{center}
\end{figure}
Let $A_{i,j}$ be the adjacency matrix of $G$ and consider the corresponding homogeneous Kuramoto model. We shall show that this system has a non-zero stable fixed point if and only if the corresponding Kuramoto Partition problem is satisfied. To begin we shall assume that we have a solution $S$ to Kuramoto Partition. Let the angles $\gamma, \alpha_{i}, \beta_{i}, i=1,..,3n$ be defined by 
\begin{eqnarray}
\pi \geq \gamma&=&2\textrm{arcos}(\epsilon)\\
\alpha_{i}&=&\textrm{arsin} \left( \sqrt{\frac{b_{i}^2 (1-\epsilon^2)}{1+b_{i}^2+2b_{i}\epsilon}} \right), i \in S\\
&=&\textrm{arsin} \left( \sqrt{\frac{b_{i}^2 (1-\epsilon^2)}{1+b_{i}^2-2b_{i}\epsilon}} \right), i \in S^c\\
\beta_{i}&=&\frac{\gamma}{2}-\alpha_{i},  i \in S\\
&=&\pi-\frac{\gamma}{2}-\alpha_{i},  i \in S^c
\end{eqnarray}

Consider the assignment of phase angles as depicted in the phase angle diagram figure 6 where for $i \in S$ $u_{i}$ is located at $\alpha_{i}, v_{i}$ is located at $\alpha_{i}+2\beta_{i}$ and all the nodes of the adjacent clique are co-located at $\alpha_{i}+\beta_{i}$, where $\sin\alpha_{i}=b_{i}\sin\beta_{i}$. For $i \in S^{c}$ $u_{i}$ is located at $2\pi-\alpha_{i}, v_{i}$ is located at $2\pi-\alpha_{i}-2\beta_{i}$ and all the nodes of the adjacent clique are co-located at $2\pi-\alpha_{i}-\beta_{i}$. It is a simple matter to verify that
\begin{eqnarray}
\sin(\alpha_{i})&=&b_{i}\sin(\beta_{i}), \\
\gamma&=&2\alpha_{i}+2\beta_{i} \textrm{ for }  i \in S\\
\sin(\alpha_{i})&=&b_{i}\sin(\beta_{i}), \\
2\pi-\gamma&=&2\alpha_{i}+2\beta_{i} \textrm{ for } i \in S^c\\
\sum_{i\in S}  \sin ({\alpha}_{i})&=&\sum_{i\in{S^c}} \sin ({\alpha}_{i})\\
\cos(\frac{\gamma}{2}) &<& \frac{1}{n}\\
\gamma &\leq& \pi
\end{eqnarray}
where (30) follows from (17) and (31) follows from (18). Equations (26)-(30) ensure that this system is a fixed point solution and (31) ensures that each $\alpha_{i}<\frac{\pi}{2}$. We must therefore have a non-zero stable fixed point solution.

\begin{figure}
\begin{center}
\includegraphics[width=70mm]{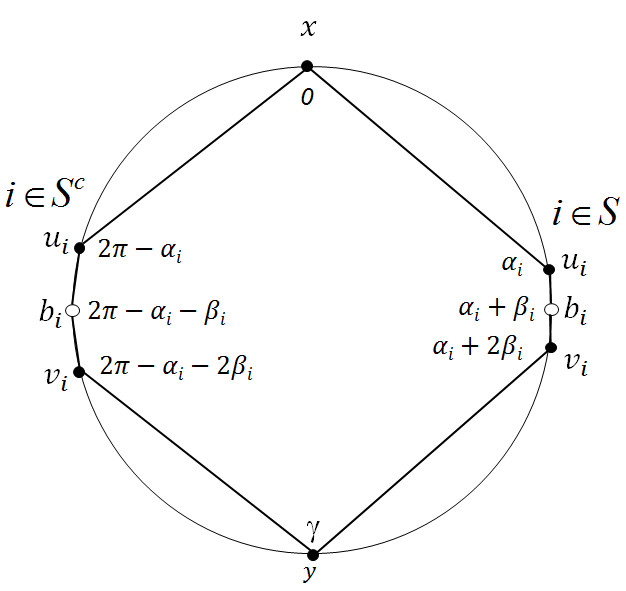}
\caption{\label{figure6.png} The phase angle structure of any non-zero fixed point solution}
\end{center}
\end{figure}

We now assume \{${\theta}^{*}_i,i=1,..,n$\} is any non-zero stable fixed point of the corresponding Kuramoto homogeneous model, and show that Kuramoto Partition must be satisfied. We do this by showing in stages that the only possible phase angle geometry follows the structure of figure 6 and that this in turn leads to a solution to  (17)-(18). First we show for each $k$ that all of the phase angles of the nodes $C_{k}$ of the clique adjacent to $u_{k}$ and $v_{k}$ must all have the same phase angles. Let $D_{k}= C_{k} \bigcup \{u_{k}\} \bigcup \{v_{k}\}$. Let

\begin{equation}
p=\sum_{j\in D_{k}} z_{j}=\sum_{j\in D_{k}}e^{i{\theta}^{*}_j}=\sum_{j\in D_{k}}(\cos({\theta}^{*}_j)+i\sin({\theta}^{*}_j)).\nonumber
\end{equation}
Then if $V$ is the complete set of all the vertices in $G$ for any $m \in C_{k}$
\begin{eqnarray}
\bar{z}_{m}p&=&\bar{z}_{m}\sum_{j\in D_{k}} z_{j}=\sum_{j\in D_{k}}\bar{z}_{m} z_{j}=1+\sum_{i\in D_{k}, j \neq m}\bar{z}_{m} z_{j}\nonumber\\ 
&=&1+\sum_{j\in D_{k}, j \neq m}\cos ({\Delta}^{*}_{jm})+i\sum_{j\in D_{k}, j \neq m}\sin ({\Delta}^{*}_{jm})\nonumber\\ 
&=&1+\sum_{j\in V}A_{jm}\cos ({\Delta}^{*}_{jm})+i\sum_{j\in V}A_{jm}\sin ({\Delta}^{*}_{jm})\nonumber\\ 
&=&1+\sum_{j\in V}A_{jm}\cos ({\Delta}^{*}_{jm}).
\end{eqnarray}
We consider two cases. If $p \neq 0$ then for some angle $\theta^p$, $p=||p||[cos\theta^p+isin\theta^p]$. Thus
\begin{equation}
\bar{z}_{m}p=||p||[\cos(\theta^p-{\theta}^{*}_m)+i\sin(\theta^p-{\theta}^{*}_m)].
\end{equation}
Comparing equations (33) and (34) we have $\sin(\theta^p-{\theta}^{*}_m)=0$. Also by the stability condition (10) $\cos(\theta^p-{\theta}^{*}_m)>0$. It follows that ${\theta}^{*}_m=\theta^p$. If on the other hand $p=0$ then $\sum_{i\in V}A_{im}\cos ({\Delta}^{*}_{im})=-1$ which contradicts the stability condition (10). Thus the nodes of every clique in $G$ must also have the same phase angle. In order to relate the non-zero stable phase angles to nodes of $G$ we let $x$ be at angle $\theta^*[x]=0$, $y$ at $\theta^*[y]$, $u_{i}$ at $\theta^*[u_{i}]$,  $v_{i}$ at $\theta^*[v_{i}]$, and the clique of nodes $C_{i}$ adjacent to $u_{i}$ and $v_{i}$ at $\theta^*[C_{i}]$. For each $i$ by the fixed point equation (9) applied with $X$=$C_{i}$ we have
\begin{equation}
b_{i}\sin(\theta^*[v_{i}]-\theta^*[C_{i}])=b_{i}\sin(\theta^*[C_{i}]-\theta^*[u_{i}])=b_{i}d
\end{equation}
for some $d$ between $0$ and $1$. This means that $\sin(\theta^*[v_{i}]-\theta^*[C_{i}])$ and $\sin(\theta^*[C_{i}]-\theta^*[u_{i}])$ are either $\textrm{arsin}(d)$ or $\pi-\textrm{arsin}(d)$. This latter possibility is ruled out however by the stability condition (10) applied at $X$=$C_{i}$. Thus 
\begin{equation}
\sin(\theta^*[v_{i}]-\theta^*[C_{i}])=\sin(\theta^*[C_{i}]-\theta^*[u_{i}]) =\textrm{arsin}(d). 
\end{equation}
Similarly by choosing $X$=$C_{i} \bigcup \{u_{i}\} \bigcup \{v_{i}\}$ then
\begin{equation}
\sin(\theta^*[y]-\theta^*[v_{i}])=\sin(\theta^*[u_{i}]-\theta^*[x]) \leq \frac{\pi}{2}.
\end{equation}
The stability condition (10) can also be used to show for each $i$ that the phase angles of $u_{i}$, $v_{i}$ and $C_{i}$ must all be on the same side of figure 6 in the sense that  $\theta^*[u_{i}]$, $\theta^*[v_{i}]$ and $\theta^*[C_{i}]$ are between $0$ and $\theta^*[y]$, or between $\theta^*[y]$ and $2\pi$. Thus the phase angle geometry must follow the structure of figure 6 where we assume by symmetry that $\gamma \leq \pi$. The fixed point conditions then take the form of equations (26)-(30). Finally we note that $n+1,..,3n$ cannot all be in $S$ or all in $S^c$. If for example $n+1,..,3n \in S$ then for $j\in S^c$ the phase angle differences between $y$ and $v_{j}$, and between $u_{j}$ and $x$ can be shown to be greater than $\frac{\pi}{2}$ so that the stability inequality (10) is violated by taking $X=C_{j} \bigcup \{u_{j}\} \bigcup \{v_{j}\}$.  The stability condition is therefore equivalent to the condition $\alpha_{j} \leq \frac{\pi}{2}$. By equation (28) and (29) this is satisfied if and only if $\gamma$ is at least $\gamma'$ where

\begin{equation}
1=\sin \left( \frac{\pi}{2} \right)=n\sin \left( \frac{\pi}{2}-\frac{\gamma'}{2}\right)=n\cos \left( \frac{\gamma'}{2}\right).
\end{equation}

Thus the stability condition is equivalent to inequality (31). We have now established the conditions (26)-(32). By setting $\epsilon=\cos(\frac{\gamma}{2})$ equation (17) follows by substitution from (26)-(29) into (30) while (18) follows from (31) and (32). We therefore have a solution to Kuramoto Partition. Thus we have shown that Kuramoto Partition is satisfied if and only if there is a  non-zero stable fixed point for the corresponding unweighted homogeneous Kuramoto model.
\subsection{Surd Partition}
The essence of the difficulty of Kuramoto Partition lies in the use of square roots with typically infinite decimal expansions. To emphasise this we provide a more generic but related partition problem that may be of interest to complexity and number theorists and conjecture that this problem is NP-hard. We note that if the square roots are omitted in the following then the problem is well known to have a polynomial solution  \cite{Garey1979}. \\
\textbf{Surd Partition}\\
Instance: A positive integer $n$ and a collection of positive integers $b_{i}, i=1,..,n$ such that $1 \leq b_{1}\leq b_{2}\leq..\leq b_{n}\leq n.$  \\
Question: Is there a subset $S$ of $\{1,2,..,n\}$ in which
\begin{eqnarray}
\left| \sum_{i\in S}  \sqrt{b_{i}} - \sum_{i\in S^c}  \sqrt{b_{i}} \right| < 1?
\end{eqnarray} 

\section{Conclusion}
In this paper we demonstrated the fundamentally complex relationship between the network topology and the solution space of the Kuramoto model by showing that determining the possibility of multiple stable fixed points from the network topology is NP-hard for the weighted Kuramoto model. Specifically we show that the NP-complete Partition problem is satisfied if and only if there is a non-zero stable fixed point solution to a related weighted homogeneous Kuramoto model.  In the case of the unweighted Kuramoto model we show that a particular real partition problem is satisfied if and only if there is a non-zero stable fixed point solution to a related unweighted homogeneous Kuramoto model.  A simplified variant of this partition problem that may be of interest to complexity and number theorists is given and conjectured to be NP-hard.  As a consequence ee conclude that it is unlikely that stable fixed points of the Kuramoto model can be characterized in terms of easily computable network invariants.
\\



\end{document}